# Exploring and Simulating Chaotic Advection:

# A Difference Equations Approach

## C. H. Skiadas

*Technical University of Crete, Chania, Crete, Greece*

**Abstract:** This paper explores the chaotic properties of an advection system expressed in difference equations form. In the beginning the Aref's blinking vortex system is examined. Then several new lines are explored related to the sink problem (one central sink, two symmetric sinks, eccentric sink and others). Chaotic forms with or without space contraction are presented, analyzed and simulated. Several chaotic objects are formulated especially when special rotation angles or a complex sinus rotation angle are introduced in the rotation-translation difference equations. Very interesting chaotic forms arise when elliptic rotation-translation equations are applied. The simulated chaotic images and attractors express several vortex-like forms resulting in various situations and especially in fluid dynamics.

# 1. Introduction

Questions addressed when dealing with chaotic advection turn back to nineteenth century and the development of Hydrodynamics and especially the introduction of the Navier-Stokes equations (Claude Navier, 1821 and George Stokes, 1845). The vortex flow case and the related forms including vortex-lines and filaments, vortex rings, vortex pair and vortex systems can be found in the classical book by Horace Lamb first edited in 1879[6]. However, the formulation of a theory that partially explains the vortex problem and gives results that coincide with the real life situations is only a matter of recent years, along with the use of computer experiments. The introduction of terms like chaotic advection and the blinking vortex system came only last decades in order to define and analyze specific vortex flow cases. In most cases the problem setting and solution followed the differential equations approach which mostly was directed to solve a boundary value problem of a Navier-Stokes equation formulation. Few interesting cases are based on a difference equations analogue in the direction to simply explain in more details the vortex flow problem. However, the formulation and analysis of vortex flow problems by using the difference equations analogue can be very useful for several cases if a systematic study is applied. In this paper we follow the difference equations methodology by introducing rotation-translation difference equations and a non-linear rotating angle along with a space contraction parameter in order to study chaotic advection problems. The interconnections between the difference and the differential equations case is also studied in specific cases.

## 2. The Sink Problem
### 2.1. *Central sink*

Consider a circular bath with a sink in the center at $(x, y) = (0,0)$. The water inside the bath is rotating counterclockwise. A colored fluid is injected in the periphery of the bath. Find the shape of the fluid filaments if the sink is open. Geometrically the problem is that of rotation with contraction following a parameter $b<1$. The rotation-translation model is applied with the translation parameter $a=0$. The equations of flow are:

$$x_{t+1} = b(x_t \cos\phi_t - y_t \sin\phi_t)$$
$$y_{t+1} = b(x_t \sin\phi_t + y_t \cos\phi_t)$$

The contraction to the radial ($r = \sqrt{(x^2 + y^2)}$) direction is found from the last relation and the equations of flow

$$r_{t+1} = b\sqrt{(x_t^2 + y_t^2)} = br_t$$

The rotation angle is assumed to follow a function of the form

$$\phi_t = c + d/r_t^2$$

The space contraction is given by estimating the Jacobian of the flow $J = b^2$. When $b<1$ a particle is moving from the periphery of the bath to the sink in the center of coordinates following spirals as is illustrated in Figure 1. The parameters selected are $b = 0.85, c = 0, d = 0.4$ and the initial point is at $(x, y) = (1,0)$.

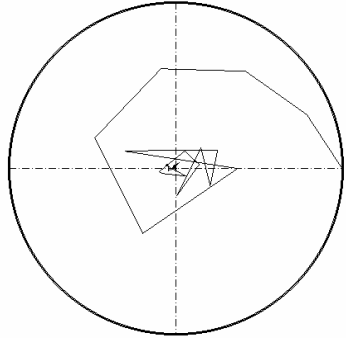

Figure 1. Spiral particle paths

When the same case is simulated for particles entering from the periphery of the rotating system at time $t = 0,1,2,...$, the following Figure 2 results.

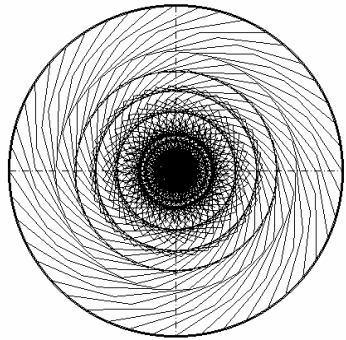

Figure 2. Spiral forms directed to the sink

The spiral forms start from the periphery and are directed toward the central sink. It is also interesting that while the spiraling flow continues, colored co centric circles appear. These circles have smaller diameter or disappear, as the rotation parameter $d$ is smaller. The parameter $b$ also influences the spiral. Next Figure 3 illustrates an advection case for parameters $b = 0.95, c = 0$ and $d = 0.01$.

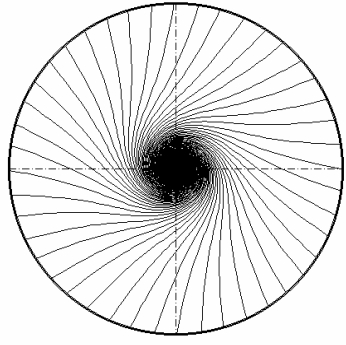

Figure 3. Spiral formation toward a central sink

## 2.2. *The contraction process*

From the above rotation-contraction equations and the very simple relation

$$r_{t+1} = br_t$$

follows that the radial contraction is

$$\Delta r_t = r_{t+1} - r_t = br_t - r_t = -(1-b)r_t$$

The differential equation for the contraction process is found by observing that

$$\frac{dr}{dt} \approx \frac{\Delta r}{\Delta t} = \frac{r_{t+1} - r_t}{(t+1) - t} = -(1-b)r$$

The resulting differential equation expressing the radial speed

$$\dot{r} = -(1-b)r$$

is solved to give

$$r = r_0 e^{-(1-b)t}$$

where $r_0$ is the initial radius.

As the equation for the rotation angle is given earlier, the movement is totally explained. The paths are spiraling toward the center. When the movement covers a full circle the new radius will be

$$r = r_0 e^{-(1-b)2\pi/\phi}$$

## 3. Eccentric Sink

In the following the case of a circular bath with an eccentric sink is analyzed. The sink is located at $(x, y) = (a, 0)$. The equations of flow are:

$$x_{t+1} = b((x_t - a)\cos\phi_t - y_t \sin\phi_t)$$
$$y_{t+1} = b((x_t - a)\sin\phi_t + y_t \cos\phi_t)$$

The rotation angle is assumed to follow an equation of the form

$$\phi_t = c + d/r_t^2$$

where $r_t = \sqrt{(x_t - a)^2 + y_t^2}$

The limit argument

$$(x_{t+1}, y_{t+1}) = (x_t, y_t) = (x, y)$$

will give the relation

$$x^2 + y^2 = b^2((x-a)^2 + y^2)$$

or after transformation

$$\left(x + \frac{ab^2}{1-b^2}\right)^2 + y^2 = \left(\frac{ab}{1-b^2}\right)^2$$

This is the equation of a circle with radius $R = \frac{ab}{1-b^2}$ centered at

$$(x, y) = \left(\frac{ab^2}{1-b^2}, 0\right)$$

The flow is not symmetric. The colored fluid starting from the outer periphery of the bath approaches the sink in few time periods as is illustrated in Figure 4. The parameters selected are $a = 0.15, b = 0.85, c = 0$ and $d = 0.1$. To simplify the process it is assumed that the colored fluid is introduced simultaneously in the periphery of the bath. Then gradually the circular form of the original colored line is transformed to a chaotic attractor located at the sink's center $(x, y) = (a, 0)$. The attractor is quite stable in form and location. Changes are possible by changing the parameter values.

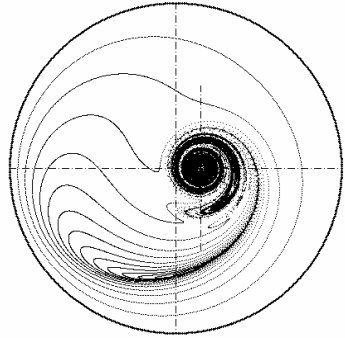

Figure 4. Chaotic attractor in eccentric sink

The attractor also appears even if the colored particles are introduced into a small region of the bath as is presented in the next Figure 5. The colored particles are introduced in a square region $(0.1*0.1)$ at the right end of the bath at $(x, y) = (1, 0)$. The parameters selected are $a = 0.15, b = 0.85, c = 0$ and $d = 0.8$. As the vortex parameter $d$ is higher than the previous case the chaotic attractor appears at the 6th time step of the process. The attractor is also larger than the previous case.

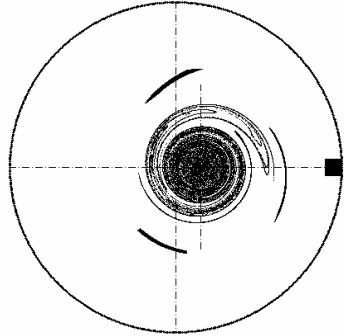

Figure 5. Chaotic attractor in eccentric sink

## 4. Two Symmetric Sinks
### 4.1. Aref's blinking vortex system
Chaotic mixing in open flows is usually modeled by using the `blinking vortex-sink system' invented by Aref,1983[1], 1984[2] and Aref and Balachandar,1986[3]. Aref's system models the out-flow from a large bath tub with two sinks that are opened in an alternating manner in order to take place a chaotic mixing in the course of the process. To model the velocity field due to a sink we assume the superposition of the potential flows of a point sink and a point vortex. If $z = x + iy$ is the complex coordinate in the plane of flow the complex potential for a sinking vortex point is

$$w(z) = -(Q + iK)\ln|z - z_s|$$

where, $z_s = (\pm a, 0)$ and $2\pi Q$ is the sink strength and $2\pi K$ the vortex strength. The imaginary part of $w(z)$ is the stream function

$$\Psi = -K \ln r - Q\phi$$

And the streamlines are logarithmic spirals defined by the function

$$\phi = -(K/Q)\ln r + const.$$

The differential equations of motion in polar coordinates are

$$\dot{r} = -Q/r$$
$$r\dot{\phi} = K/r$$

And their solutions are

$$r = \sqrt{r_0^2 - 2Qt}$$

and

$$\phi = \phi_0 - (K/Q)\ln(r_t/r_0)$$

The flow of the system is fully characterized by the non-dimensional sink strength $\eta = QT/a^2$ and the ratio of vortex to sink strength $\xi = K/Q$. $T$ is the flow period and $a$ is the distance of each sink from the center of coordinates. As it is indicated in the literature (Károlyi and Tél,1997[5], Károlyi et al,2002[4]) chaotic flow appears for parameter values $\eta = 0.5$ or larger and $\xi = 10$. More precisely when particles are injected into the flow in few time periods are attracted in a specific region (the attractor) of the flow system. Several studies appear last years investigating the phenomenon theoretically and experimentally. The theoretical studies include also simulations by using large grids (1000x1000) and arithmetic solution of the general equations of flow. These studies suggest that the attractors are time periodic according to the time periodicity of the flow. However, if only one sink is used a stable attractor could be present at least theoretically and following simulation experiments as is presented above. This is modeled by investigating the geometry of the flow. First of all Aref's blinking vortex system is applied in a rotating fluid. We select a counter-clockwise rotation. The symmetric sinks are located at $(x, y) = (-a, 0)$ and $(x, y) = (a, 0)$ and the time period is $T = 1$. According to this system the flow is not stationary and there are jumps in the velocity field at each half period $T/2$. In other words a particle located at $(-a, 0)$ appears at $(a, 0)$ the next time period as is illustrated in the next graph of Figure 6.

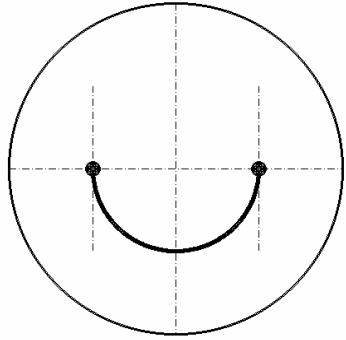

Figure 6. The two symmetric sinks model

The modeling we propose is to analyze a discrete time system, as is Aref's system by using the theory of difference equations and discrete systems. It looks more convenient and highly simpler considering the geometry of this system. The model we search must be a rotation-translation one with a parameter $b < 1$ expressing the gradual shortening of the radius $r$, which leads the particles to follow logarithmic spiral trajectories around the sinking vortices. Following to above theory a rotation-translation model of this type is expressed by the difference equation:

$$z_{t+1} = a + b(z_t - z_s)e^{i\phi_t}$$

The above complex equation can be written as

$$x_{t+1} + iy_{t+1} = a + b[(x_t + a) + iy_t](\cos\phi_t + i\sin\phi_t)$$

The system of iterative difference relations for $x$ and $y$ is obtained by equating from both sides the real and the imaginary parts of the last complex formula

$$\begin{aligned} x_{t+1} &= a + b[(x_t + a)\cos\phi_t - y_t \sin\phi_t] \\ y_{t+1} &= b[(x_t + a)\sin\phi_t + y_t \cos\phi_t] \end{aligned}$$

If a particle is located at position $(x, y) = (-a, 0)$ the next point after time $t = 1$ will be located at $(x_1, y_1) = (a, 0)$. The next problem we have is to define the form of the function of the angle $\phi$. From the original differential equations of flow the differential equation for $\phi$ is

$$\dot\phi = QK/r^2$$

where $r = \sqrt{(x+a)^2 + y^2}$

As the value of $\Delta t$ is equal to the periodic time $T = 1$ the last equation for $\dot\phi$ yields:

$$\dot\phi = \frac{d\phi}{dt} \approx \frac{\Delta\phi}{\Delta t} = \frac{\phi - \phi_0}{T} = \phi - \phi_0$$

And thus

$$\phi = \phi_0 + \frac{\eta\xi}{r^2}$$

Now it is very easy to find that for Aref's blinking vortex flow the value of $\phi_0 = \pi$ in order to have a half-cycle rotation from one sink to the other. The last equation for $\phi$ is also written as

$$\phi = \phi_0 + \frac{d}{r^2} = c + \frac{d}{r^2}$$

Where $d = \eta\xi$ is the vortex strength. For the experiments presented in the literature $\eta = 0.5, \xi = 10$ and thus $d = 5$. However, the chaotic region is more wide as is illustrated in the next figures.

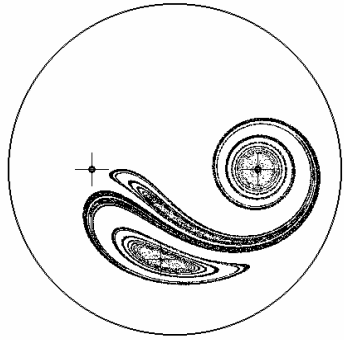

Figure 7. Chaotic attractor in the two-sink problem

The graph of Figure 7 of a chaotic attractor illustrates the two-sink case for parameter values $a = 1, b = 0.8, c = \pi$ and $d = 3$. There are two main vortex forms counter-balancing each other. The first form is located at the right hand side sink at $(x, y) = (a, 0)$. The second vortex form is centered at $(x, y) = (a + 2ab\cos(\phi), 2ab\sin(\phi))$, where $\phi = d/(4a^2)$. The two main vortex forms can be separated when the parameter $d$ expressing the vortex strength is relatively small. Such a case is presented in the next Figure 8. The parameter $d = 1$ while the other parameters remain the same with the previous example. The attractor is now completely separated into two chaotic vortex forms (attractors).

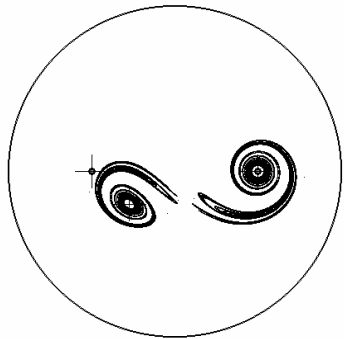

Figure 8. Two distinct vortex forms ($d = 1$)

Another idea is to give high values to the parameter $d$ expressing the vortex strength. The selection of a value $d = 2\pi$ for the vortex strength parameter leads to a more complicated vortex form as is presented in the next Figure 9. There are three equilibrium points for time $t = 1,2,3$. The first of these points is the center of the right hand side sink.

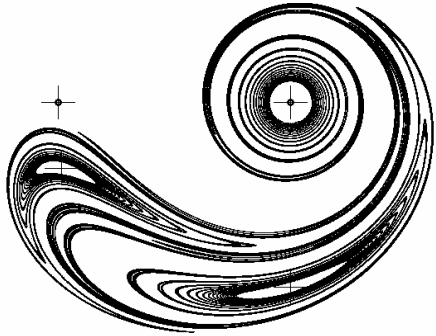

Figure 9. The chaotic attractor with strong vortex strength parameter $d = 2\pi$

Another point is to find the vortex form in the case of three sinks located in an equilateral triangle. The simulation of this situation is achieved by selecting a value $c = 2\pi/3$ for the sink strength. The other parameters are $b = 0.95, d = \pi$ and $a = 5$. Figure 10 illustrates this case.

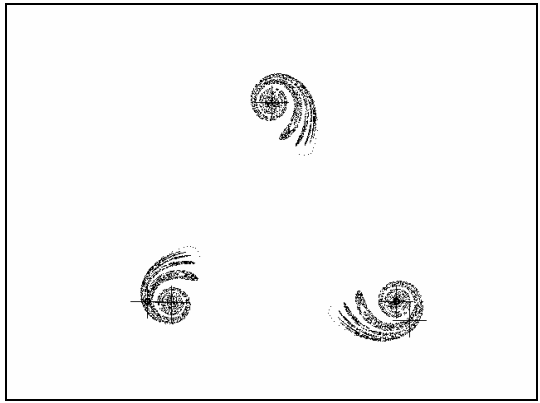

Figure 10. Three vortex forms

A four vortex form is presented in the next Figure 11. This is achieved by assuming a special value $c = 2\pi/4$ for the parameter $c$, thus dividing the total circle in four sectors. The positions of the four sinks are located on the corners of a square. The parameters selected for the simulation are $b = 0.98, d = \pi$ and $a = 5$.

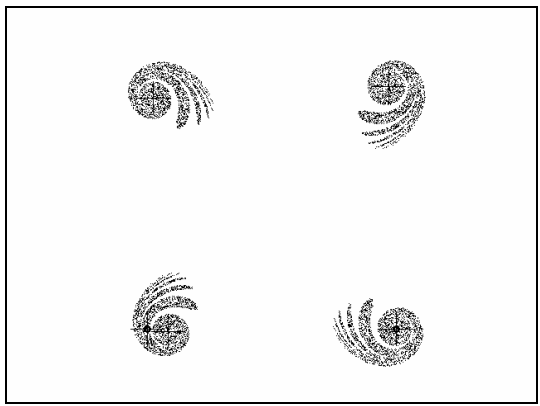

Figure 11. Fourth order vortex forms

As before, but for $c = 2\pi/5$, whereas the other parameters remain unchanged, a fifth order vortex form results illustrated in Figure 12.

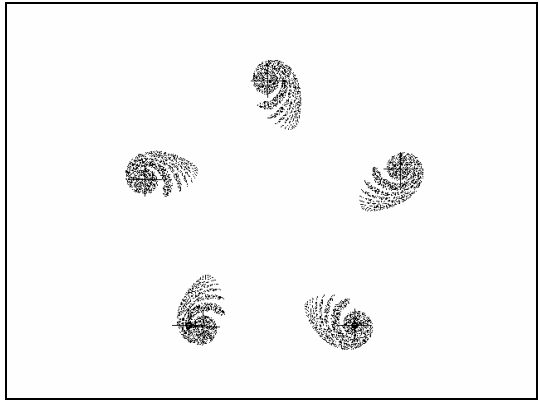

Figure 12. Five vortex forms

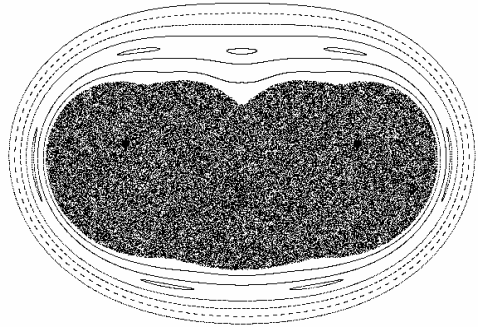

Figure 13. One chaotic image ($d = 1$)

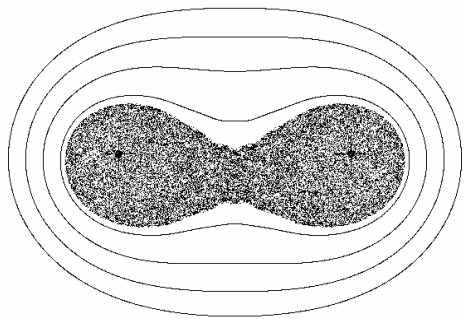

Figure 14. Starting division into two images ($d = 0.4$)

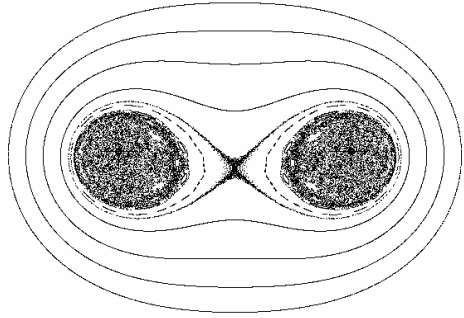

Figure 15. Division into two images ($d = 0.3$)

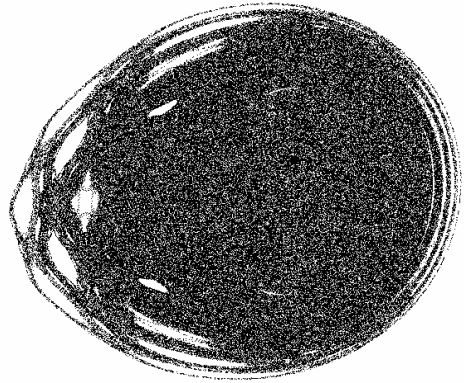

Figure 16. Enlargement of the previous right hand side attractor

## 5. Chaotic Forms Without Space Contraction

Chaotic forms produced earlier in this paper but without space contraction, by means when $b = 1$ and having Jacobian $J = 1$, is a special case of chaotic modeling. As there is no radial or space contraction the flow could be chaotic but the resulting chaotic attractors are not classified as main vortex forms. The following example, Figure13, illustrates the case with parameters $a = 1, b = 1, c = \pi$ and $d = 1$. The chaotic form is symmetric. The axis $y = 0$ is the symmetry axis. A division of the main chaotic form is starting. As the vortex strength parameter $d$ is decreasing ($d = 0.4$ for Figure 14 and $d = 0.3$ for Figure 15) two distinct symmetric chaotic forms arise. The last Figure 16 gives an enlargement of the right hand side chaotic attractor produced when $d = 0.3$. Interesting chaotic details are present in this attractor.

## 6. Other Chaotic Forms

In this computer experiment we use the classical rotation-translation model with a rotation angle expressed by an inverse function of the square of the radius $r$.

$$\theta_n = c + \frac{d}{x_n^2 + y_n^2}$$

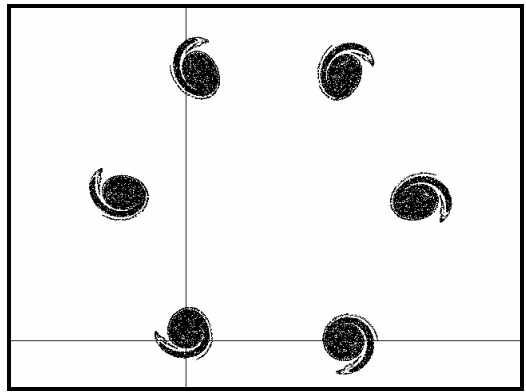

Figure 17. Sixth order vortex forms

The task is to vary the rotation parameter $c$ while the other parameters remain stable. The space contracting parameter $b$ takes a value close to unity ($b = 0.99$), which could express speeds of low level directed toward the sink. When $a = 10$, $d = 10$ and $c = 1$ six rotating vortex-like spiral objects appear in Figure 17.

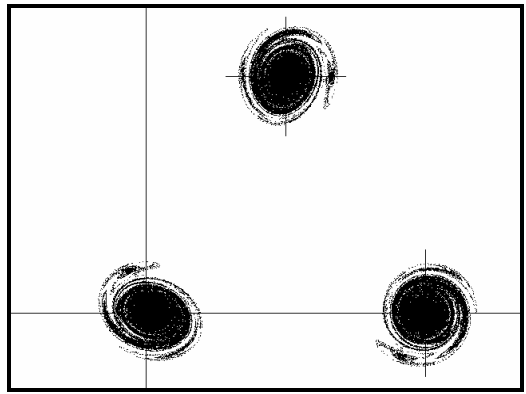

Image 18. Third order vortex forms

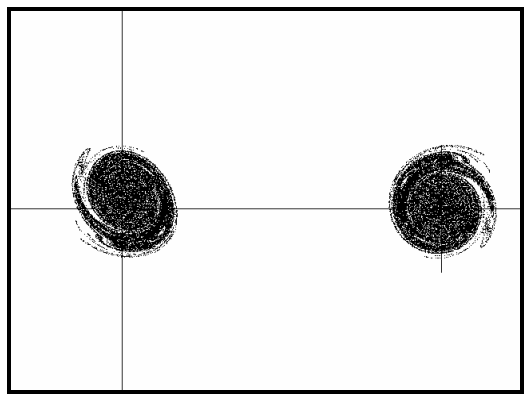

Image 19. Second order vortex forms

When the parameter $c = 2$, while the other parameters remain unchanged three objects appear illustrated in Figure 18. These chaotic forms illustrate two-armed spiral galaxy-like objects located at $(x, y) = (0,0)$, $(x, y) = (a,0)$ and $(x, y) = (a + ba\cos(ta), ba\sin(ta))$ where $\theta_a = c + d/a^2$.

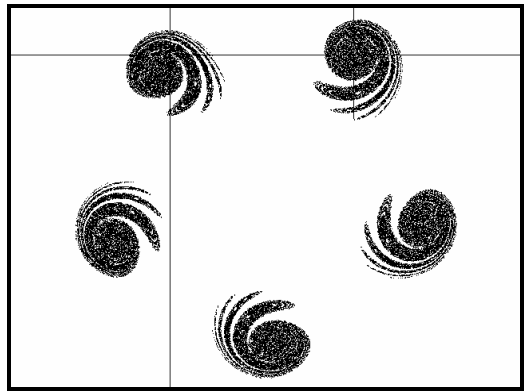

Figure 20. Fifth order vortex forms

The chaotic form of the attractor is turned to be merely two one-armed spirals for $c = 3$ as is presented in Figure 19. The chaotic attractors are located at $(x, y) = (0,0)$ and $(x, y) = (a,0)$. Five chaotic attractors appear when the parameter $c = 5$ presented in Figure 20.

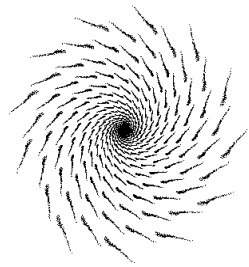

Figure 21. Spiraling toward the sink ($R = 1$)

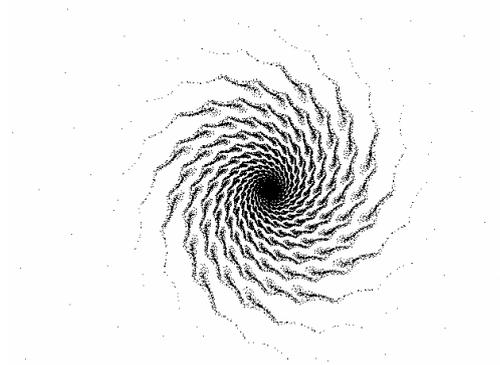

Figure 22. Spiraling toward the sink ($R = 2$)

When $c = 0.5$ a multi-armed spiral is present illustrated in Figure 21. The original rotating disk has radius $R = 1$. When $R = 2$ the following Figure 22 appears.

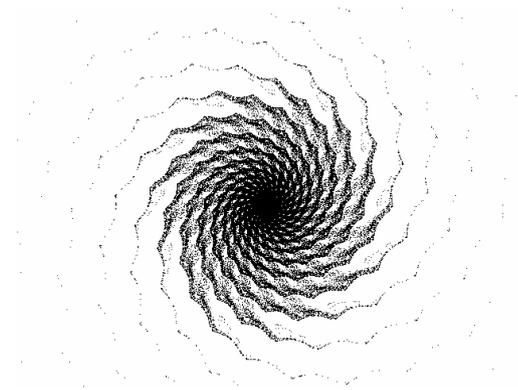

Figure 23. Spiraling toward the sink ($R = 3$)

The number of spirals remains the same, but more details are present in Figure 23. The original rotating disk has radius $R = 3$.

## 7. Complex Sinus Rotation Angle

Very interesting chaotic attractors arise by assuming that the rotation angle of the rotation-translation model is a complex sinus function. For the first application the following function for the rotation angle is proposed

$$\theta_n = \frac{1}{\sin(c - d/(1 + x_n^2 + y_n^2))}$$

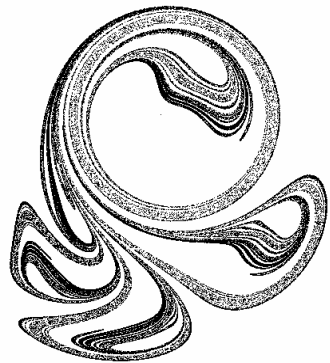

Figure 24. A sinus rotation angle chaotic attractor ($c = 0.4, d = 6$)

The set of parameters is $a = 1$ for the translation parameter, $b = 0.83$ for the space contraction parameter and $c = 0.4$ and $d = 6$ for the rotation parameters. Figure 24 illustrates the resulting chaotic attractor. The image shows a circular ring along with an internal vortex form and an external part formed from three main vortex forms and a smaller fourth one.

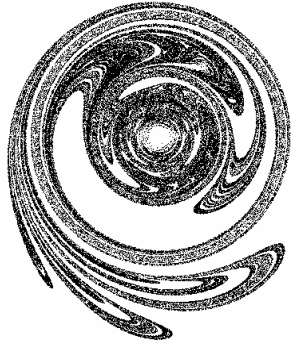

Figure 25. A sinus rotation angle chaotic attractor $\theta_n = 1/\sin(c - d/r_n)$

A change in the function for the rotation angle will give another (Figure 25) chaotic rotation image. The parameters remain the same as in the preceding example while the rotation angle function is

$$\theta_n = \frac{1}{\sin(c - d/\sqrt{x_n^2 + y_n^2})} = \frac{1}{\sin(c - d/r_n)}$$

The resulting image has a circular ring with two main vortex forms and a small one in the outside region and vortex forms inside interconnected with a central rotating part.

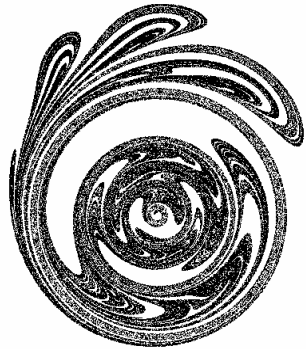

Figure 26. A sinus rotation angle chaotic attractor ($d = 16$)

A relatively simpler function for the rotation angle as is the

$$\theta_n = \frac{1}{\sin(d/(1 + x_n^2 + y_n^2))}$$

has only one parameter. However the resulting chaotic image in Figure 26 is also quite complicated. Again there is a circular ring and three outside of the ring connected vortex forms while several vortex forms rotate clockwise and counterclockwise inside the ring. The parameters selected are $a = 1.2$ for translation, $b = 0.85$ for space contraction and $d = 16$ for rotation.

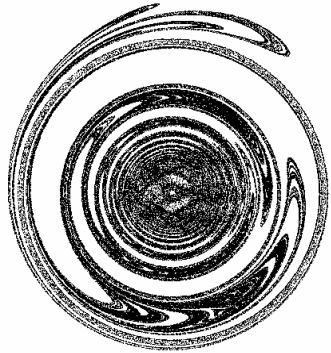

Figure 27. A sinus rotation angle chaotic attractor ($\theta = 1/\sin dr^2$, $b = 0.7$)

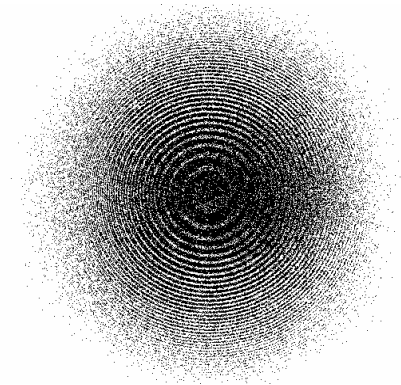

Figure 28. A sinus rotation angle rotation image ($\theta = 1/\sin dr^2$, $a = 0.85, b = 0.8$)

A completely different image results by introducing another function for the rotation angle

$$\theta_n = \frac{1}{\sin(d(x_n^2 + y_n^2))} = \frac{1}{\sin(dr^2)}$$

The same rotation-translation case with parameters $a = 1.2$, $b = 0.85$ and $d = 16$ provides also a circular ring image (Figure 27) with surrounding vortex forms, but now, the outside the ring vortex forms are small whereas the inside part is larger including a central circular bulge and vortex forms rotating clockwise and counterclockwise. The standard rotation-translation model is also applied with parameters $a = 1.1$, $b = 0.8$ and $d = 16$. Figure 28 illustrates this case. A large number of co-centric rings appear.

## 8. A Special Rotation-Translation Model

This rotation-translation model is given by the following set of relations

$$\begin{bmatrix} x_{n+1} \\ y_{n+1} \end{bmatrix} = \mathbf{rot}\,\theta_n \begin{bmatrix} b(x_n^2 - y_n^2) \\ 2bx_n y_n \end{bmatrix} + \begin{bmatrix} a \\ 0 \end{bmatrix} = \begin{bmatrix} a + b[(x_n^2 - y_n^2)\cos\theta_n - 2x_n y_n \sin\theta_n] \\ b[(x_n^2 - y_n^2)\sin\theta_n + 2x_n y_n \cos\theta_n] \end{bmatrix}$$

where the rotation angle is given by the relation

$$\theta_n = \frac{d}{r_n}$$

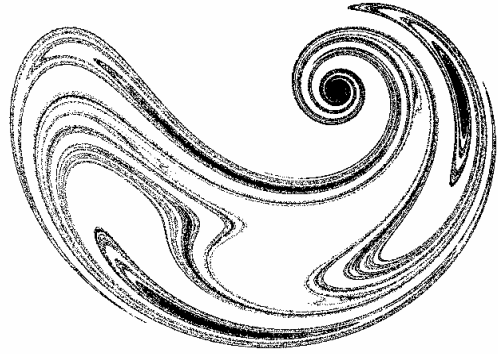

Figure 29. Chaotic image of a special rotation-translation map

This map is of special attention as the Jacobian is $J = 4b^2 r^2$. When $r = 1/(2b)$ the map is area preserving whereas, when $r < 1/(2b)$ a space contraction takes place. For $r > 1/(2b)$ area expansion takes place. The parameter values selected for the application are $a = 0.8$ for the translation parameter, $b = 0.6$ for the area contraction parameter and $d = 5$ for the rotation parameter. Figure 29 illustrates the simulation results. There is a rotation center and vortex-like forms.

## 9. Other Rotation-Translation Models
### 9.1 *Elliptic rotation-translation*
The following system of two difference equations provides elliptic forms when the ellipticity parameter $h \neq 1$.

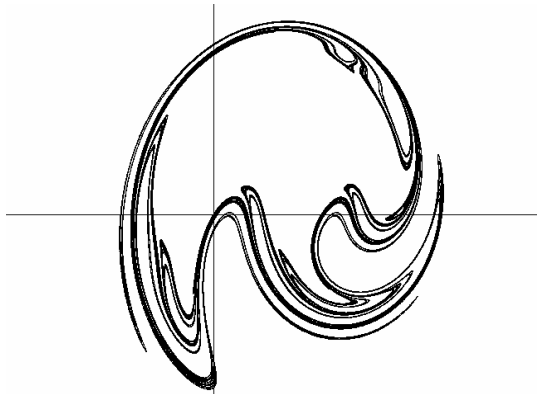

Figure 30. Chaotic image of an elliptic rotation-translation map

$$x_{n+1} = a + b(x_n \cos\theta_n - h y_n \sin\theta_n)$$
$$y_{n+1} = b(x_n \sin\theta_n / h + y_n \cos\theta_n)$$

This system results after a transformation of the classical rotation-translation system. The rotation angle is of the form $\theta = dr^2$. The parameters selected are $a = 0.5$ for translation, $b = 0.9$ for space contraction, $d = 3$ for rotation and $h = 0.8$ for ellipticity. The resulting image is illustrated in Figure 30. The elliptic rotation equations give a deformed image compared to the previous circular type cases presented in this chapter.

### 9.2 *Rotation-translation with special rotation angle*
This case is modeled by using the classical rotation-translation model with space contraction but now the rotation angle obeys the function

$$\theta_n = c + \frac{d}{r_n^2 - e}$$

where for our example $e = 4.5$ and the other parameters are $a = 3$, $b = 0.9$, $c = 1.52$ and $d = 1.014$.

A space contraction parameter $e$ is inserted in the function expressing the rotation angle. The resulting image is illustrated in Figure 31. This and the following images are of particular interest when studying chaotic advection cases. The main part of the image is a circular ring centered at $(x, y) = (a, 0)$. Three vortex forms appear outside the circular ring and two smaller vortex forms appear inside the ring. All the vortex forms are originated from the periphery of the circular ring. Smaller vortex forms appear inside every main vortex form. In the same figure the two axes of coordinates and the part of the circle expressing the space curvature due to the space contraction parameter $b$ and the translation parameter $a$ are illustrated. This circle is of radius $R = \frac{ab}{1-b^2}$ centered at $(x, y) = (\frac{a}{1-b^2}, 0)$.

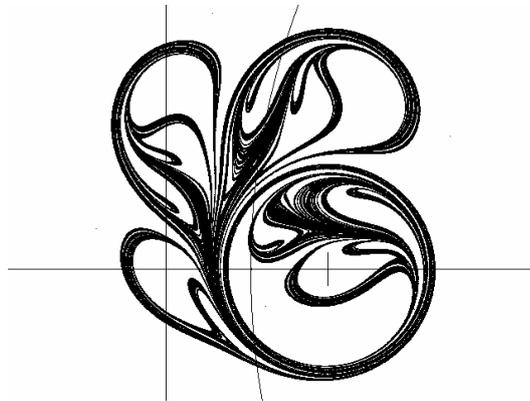

Figure 31. Chaotic image of an area contraction rotation angle ($c = 1.52$ and $d = 1.014$)

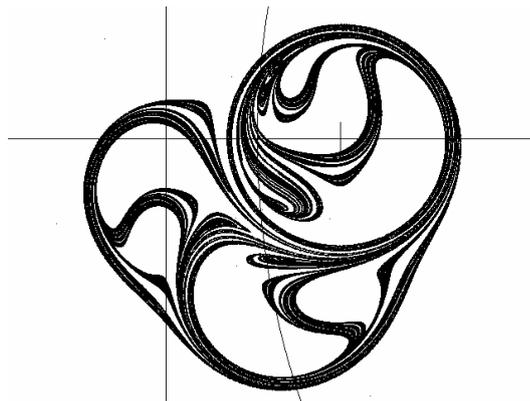

Figure 32. Chaotic image of an area contraction rotation angle ($c = 4.3$ and $d = 1$)

As the parameter $c$ is changed the resulting image shows the same circular ring located at the same center and having similar magnitude but changes appear in the vortex forms. Figure 32 expresses a similar to the previous case but now the parameters $c = 4.3$ and $d = 1$, whereas the other parameters remain unchanged. The two vortex forms inside the circular ring keep more space while the outer vortex forms form a main vortex image with vortex forms inside.

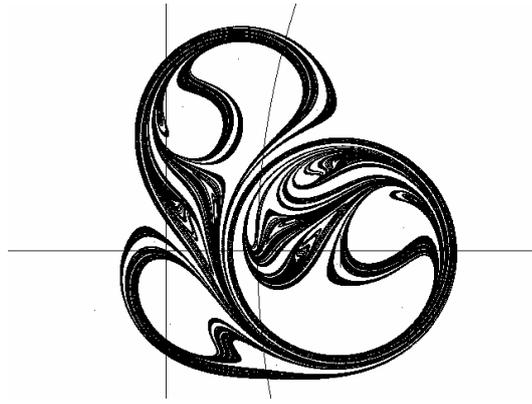

Figure 33. Chaotic image of an area contraction rotation angle ($c = 8.5$)

Figure 33 has the same parameters as the previous case but now the parameter $c = 8.5$. The outer vortex forms are two. However the image is similar to that of Figure 31.

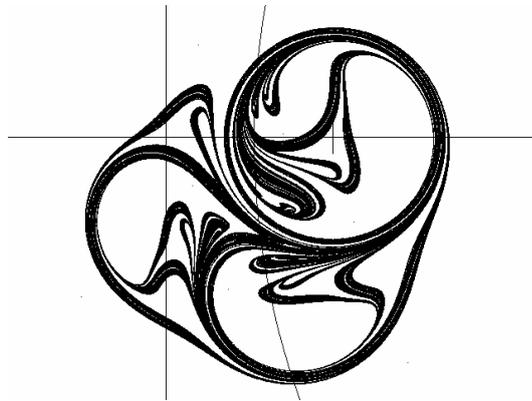

Figure 34. Chaotic image of an area contraction rotation angle ($c = 17$)

Figure 34 illustrates a chaotic image similar to the previous case but now the parameter $c = 17$. The image is similar to that in Figure 32.

## 10. Conclusions

In this paper we examined the chaotic properties of chaotic advection systems starting from the classical Aref's blinking vortex system. The study followed a difference equation methodology which is, in several cases, more simple and more instructive from the differential equations analogue. We analyzed and applied a rotation-translation set of difference equations with a dynamical non-linear rotation angle. The resulting chaotic images and attractors express several vortex-like forms resulting in various situations especially in fluid dynamics.